\documentclass[12pt]{article}
\usepackage{times}
\usepackage{amsmath}
\usepackage{amsfonts}
\usepackage{amssymb}
\usepackage{graphicx}
\usepackage{latexsym}
\usepackage{epstopdf}
\usepackage{url}

\begin{document}

\title{A Behavioural Foundation for Natural Computing and a Programmability Test\footnote{Invited Talk at the \emph{Symposium on Natural/Unconventional Computing and its Philosophical Significance}, \textit{AISB/IACAP Alan Turing World Congress 2012}.}}

\author{Hector Zenil\\
Institut d'Histoire et de Philosophie des Sciences et des Techniques\\ (Paris 1 Sorbonne-Panth\'eon/ENS Ulm/CNRS), Paris, France.\\
hector.zenil-chavez@malix.univ-paris1.fr}

\date{}

\maketitle

\begin{abstract}
What does it mean to claim that a physical or natural system computes? One answer, endorsed here, is that computing is about programming a system to behave in different ways. This paper offers an account of what it means for a physical system to compute based on this notion. It proposes a behavioural characterisation of computing in terms of a measure of programmability, which reflects a system's ability to react to external stimuli. The proposed measure of programmability is useful for classifying computers in terms of the apparent algorithmic complexity of their evolution in time. I make some specific proposals in this connection and discuss this approach in the context of other behavioural approaches, notably Turing's test of machine intelligence. I also anticipate possible objections and consider the applicability of these proposals to the task of relating abstract computation to nature-like computation.\\

\end{abstract}

Faced with the question of computation, it may be tempting to go along with the formal mathematical position and simply invoke Turing's model. This paper doesn't need to do this, though its author couldn't be more wholehearted in granting the beauty and generality of the universal Turing machine model, which, it will be argued, is also a natural foundation for unconventional (and natural) computation.

To date the study of the limits of computation has succeeded in offering us great insight into this question. The borderline between decidability and undecidability has provided an essential intuition in our search for a better understanding of computation.  One can, however, wonder just how much can be expected from such an approach, and whether other, alternative approaches to understanding computation may complement the knowledge and intuition it affords, especially in modern uses of the concept of computation, where objects or events are seen as computations in the context of physics. 

One such approach involves not the study of systems lying ``beyond'' the uncomputable limit (the ``Turing limit"), but rather systems at the farthest reaches of the computable, in other words the study of the minimum requirements for universal computation. How easy or complicated is it to assemble a machine that is Turing universal? This minimalistic bottom-up approach is epitomised by Wolfram's programme~\cite{wolfram} in its quest to study simple programs, a programme initiated by Minsky~\cite{minsky} and to which several authors have contributed (see~\cite{woods} for an excellent survey). The underlying question is how pervasive and ubiquitous the computational property of universality is in computational and natural systems. From the various results concerning small universal computing systems, we now know that generating universality takes very little, indeed that it seems to be the case that it is more difficult to design a non-trivial non-Turing-complete computer language than a Turing-complete one. Thus it seems natural to believe that computation and universality are not exclusive to digital computers. 

This paper is organised as follows. In Section~\ref{foundation}, the foundations of natural computation are discussed, taking as a starting point Turing's case---argued in relation to digital computation---for the disembodied essence of natural computation. In Section~\ref{approach}, the behavioural approach to natural computation will be introduced, based on notions of algorithmic complexity, and with an analogy drawn between it and Turing's pragmatic approach to machine intelligence. In Section~\ref{taxonomy}, a characterisation and taxonomy of computation (and of computers) based on the compression-based approximation of a system's algorithmic complexity is advanced and, finally, in Section~\ref{objections} possible objections are analysed, also in light of the way in which they can be transferred between Turing's test and the definition of nature-like computation adopted herein.

\section{A classical foundation for unconventional computation}
\label{foundation}

A compiler written between computational systems, hence a mapping between symbols and states, is the usual way of proving in a technical fashion that one system is equivalent to another in computational power (hence that \emph{it computes}). A legitimate question that arises is whether we need this technical apparatus to define computation. The problem can be phrased in the words of M. Conrad~\cite{conrad} \emph{In the real world, little if anything is known of the primitive operations or symbols of a system.}

One strong criticism of the idea that natural objects (including the universe) compute is that the question and answer become meaningless, as it is hard to see how any physical system would not be computational~\cite{putnam,searle}. One concept that Turing did not advance (although he suggested taking into account the percentage of people acknowledging the success or failure of his machine intelligence test \cite{turing}), but that is very much in the spirit of another of his seminal contributions (the relativisation of computation, in his notion of degrees of computation~\cite{turing2}), is a metric of intelligence, one where passing or failing is beside the point, but which tells us how close or far we are from intelligent behaviour. 

This paper advances a metric of approximative, asymptotic and limit behaviour, not for intelligence, but for computation, one that identifies objects to which some degree of computation can be assigned on the basis of how they behave, and particularly on the basis of whether they can be programmed. It thereby places programmability at the centre of our definition of computation and so avoids representationalism.

\subsection{A behavioural approach to computation}

\begin{figure}[htdp]
\label{turingtest}
\centering
  \scalebox{.25}{\includegraphics{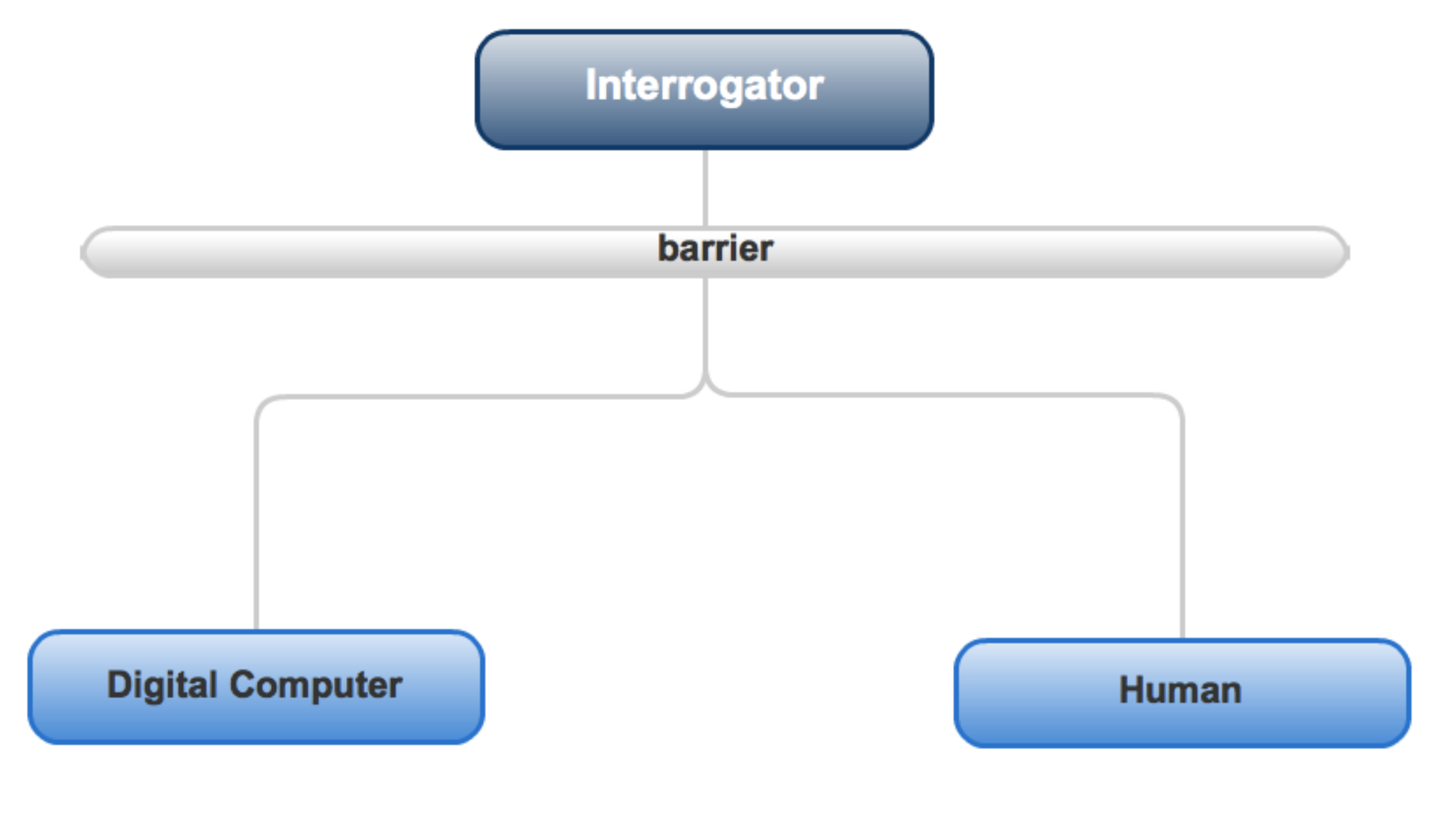}}\\
\caption{The basic elements of Turing's test of intelligence.}
\end{figure}

Among the most important of Turing's contributions to AI was his test of machine intelligence \cite{turing}, devised as a response to the question of whether computers could think. The Turing test is a pragmatic behavioural approach to the problem of assigning intelligence to objects (see Fig.~\ref{turingtest}). In the spirit of Turing, one may ask whether objects other than electronic computers compute, in particular natural objects and natural processes. This question ultimately leads to the more general question of whether the universe itself computes (also known as ``pancomputationalism"), and if so how. Some speculative answers have been given, but in this presentation we take a more pragmatic and behavioural approach to the question, in the spirit of Turing's approach to intelligence. 

When Alan Turing was thinking about AI he believed ``that in about fifty years' time it will be possible to programme computers, with a storage capacity of about $10^9$, to make them play the imitation game so well that an average interrogator will not have more than a 70 percent chance of making the right identification after five minutes of questioning. $\ldots$ I believe that at the end of the century the use of words and general educated opinion will have altered so much that one will be able to speak of machines thinking without expecting to be contradicted." 

Most would agree that Turing's faith hasn't exactly been vindicated, perhaps because of the way in which the definition of intelligence has changed over time, indeed every time that some task requiring intelligence has been successfully executed by a computing machine, from crunching numbers faster than humans to faring better at chess, and more recently, performing some rather complicated games on TV shows. I think we live in a time where it has finally become common practice to treat objects other than electronic and human computers as computing objects, and so I shall address the ineluctable generalisation of the concept of computation beyond the realm of digital computers, and more specifically its extension to natural systems. If Turing's claim were to be revised, with \emph{objects computing} being substituted for "machines thinking", the prediction seems right on target: ``I believe that at the end of the century the use of words and general educated opinion will have altered so much that one will be able to speak of [all kinds of objects computing] without expecting to be contradicted."

\subsection{Digital computation as natural computation}
\label{universality}

Turing's most important contribution to science is his definition of universal computation, integral to his attempt to mechanise the concept of a computing machine. A universal (Turing) machine is an abstract device capable of carrying out any computation for which an instruction can be written. More formally, given a fixed description of Turing machines, we say that a Turing machine $U$ is universal if for any input $s$ and Turing machine $M$, $U(\langle M\rangle, s)$ halts if $M$ halts on $s$ and outputs $M(s)$; and does not halt if $M(s)$ does not (where $\langle M \rangle$ means the codification of $M$ in bits so that it can be fed to a Turing machine $U$ that accepts binary inputs). In other words, $U$ is capable of running any Turing machine $M$ with input $s$. 

The fact that we need hardware and software is an indication that we need a programmable substratum that can be made to compute something for us, but Turing's main contribution vis-\`a-vis the concept of computational universality is that data and programs can be stored together in a single memory without any fundamental distinction. One can always write a specific-purpose machine with no input to perform any computation, and one can always write a program describing that computation as the input for a (universal) Turing machine, so in a sense there is a non-essential distinction between program and data.

It is clear that one can derive a fundamental kind of natural computation from Alan Turing's seminal concept of universal computation. Turing points out~\cite{turing} that given that Babbage's computer did not use electrical power, and that because Babbage's and all digital computers are in some fundamental sense equivalent, electricity cannot be a fundamental property of computation. Neither is it the carrier. In other words, Turing universality disembodies computation, uncoupling it from any physical substratum. This doesn't mean that one can carry out computations without physical elements, but rather that the nature of the physical elements is not very relevant except insofar as it bears upon the (important) question of resources (capacity, speed). A programmer uses memory space and cpu cycles in a regular computer to perform a computation, but this is by no means an indication that computation requires a computer (say a PC), only that it needs a substratum. The behaviour of the substratum is the underlying property that makes something a computation.

 The main difference between a digital electronic computer and a natural system that possibly computes, is that the former was designed for the purpose, and hence one can easily identify all its elements and have recourse to them when establishing a definition of computation. For natural systems, however, there is little hope that even if their elements were to be identified, one could define their states in a way that captured all their convolutions well enough to establish that they possessed some property of computation. This situation is not that different from the undecidability of the halting problem, but it is in some sense more general. For digital computation, the undecidability of the halting problem means that if one wished to know whether a computation would eventually halt, one would have no other option than to run it and wait and see (possibly for an infinite length of time). In natural systems, the halting problem is closer to the reachability problem, that is, the question of whether a system will reach a certain configuration. By reduction to the halting problem, this can also be proven to be undecidable. The halting and reachability problems are in a strong sense behavioural and subjective in nature, as the behaviour of a system has to be determined by waiting, witnessing and recording it so that it can be understood in retrospect. If for Turing machines $M$, the function that computes $M$ cannot in general be found, there is little hope of ever finding or even defining the function of a natural system. Hence one has to give up on trying to define computation for natural systems using elements such as states or functions. 

We know that systems that nobody ever designed as computers are able to perform universal computation, for example Wolfram's Rule 110 \cite{wolfram,cook} (in the rulespace of the so-called elementary cellular automata \cite{wolfram}), and that this, like other remarkably simple systems, is capable of universal computation (e.g. Conway's game of Life \cite{gol} or Langton's ant~\cite{langton}). These systems may be said to readily arise physically, not having been deliberately designed.  There is, however, no universal agreement as regards the definition of what a computer may or may not be, or as to what exactly a computation might be, even though what computation is and what a computer is are well grasped on an intuitive level. 

Now we would like a concept of computation associated with natural and physical phenomena that we can measure and build on. We want a metric of computation that allows us to identify what is a computer and what is not. We want to be able to distinguish what computes from what does not. And we want a metric that we can use.

\section{A Turing test-inspired approach to computation}
\label{approach}

As for Turing's test of intelligence, where one needs to accept that humans think if the test is to make sense, the reader must first accept that digital computation is performed in nature and that nature is capable of digital computation, even if only by the digital computers constructed by humans for precisely such a (general) purpose. Human behaviour is to the Turing test what digital computation is to this behavioural approach to natural computation. The argument can be rendered more succinctly thus: Electronic computers compute, electronic computers are physical objects, physical objects are part of the universe, a part of the universe is therefore capable of computation. Computers can be seen as the result of the re-programming of a part of the universe to make it compute what we want it to compute. This means that the question is not exactly whether the universe is capable of digital computation but rather whether the universe \emph{only} performs computation, and if so, what kind of computation. I aim to provide a behavioural definition of computation that admits a wider definition of the notion of `computation'. Notice that I am replacing the question of whether a system is capable of digital computation with the question of whether a system can behave like a digital computer and whether a digital computer can exhibit the behaviour of a  natural system. So the approach is still classical in this sense, but purposely neutral with regard to the ontological issue. Also notice again the similarity with Turing's approach to machine intelligence. Turing chose to sometimes speak of ``imitation" instead of ``behaviour". ``Imitation", however, seems to carry connotations of intentionality (see Subsection~\ref{intentionality}), and I am not very comfortable with the suggestion that a natural system may have a will to, or may purposefully imitate another system, especially if it is forced to do so artificially (although imitation is quite common in nature, where, for example, some animals mimic the behaviour of other animals to avoid being preyed upon).

 To make sense of the term ``computation'' in the contexts I'm interested in (modern views of physics), I propose a behavioural notion of nature-like computation (similar in spirit to the coinage ``physics-like computation"  \cite{margolus,sutner}) that is compatible with digital computation but meaningful in broader contexts, independent of representations and possible carriers. This will require a measure of the degree of programmability of a system based on a compressibility index which is ultimately rooted in the concept of algorithmic complexity. I ask whether two computations are the same if they look the same and I try to answer with a specific tool possessing the potential to capture a notion of qualitative behaviour.

 In \cite{cronin}, a similar approach, but this time to the question of life, is audaciously put forward, also in the spirit of Turing's test. The idea is to recognise living systems by the way they behave and communicate through the signals transmitted between biological cells. This approach uses a biological interrogator to ask not what life is but rather when an artificial cell can be said to be alive.

\begin{center}
\begin{table}[h]
   \begin{center}
   \footnotesize
      \tabcolsep=0.12cm
   \begin{tabular}{|c|c|c|c|}
\hline
&\textit{Turing test for} & & \textit{Turing test for}\\
&\textit{intelligence} \cite{turing} & \textit{Turing test for life} \cite{cronin} & \textit{computation}\\
\hline
\textbf{Imitated property} & Thought & Cellular functions & Programmability \\
\hline
\textbf{Subjects} & Computing & Biological and artificial & any object\\
\textbf{in question}& machines &  cells &\\
\hline
\textbf{Embodiment of} & Human & Biological life & Digital \\
\textbf{property} & intelligence & (metabolism, evolution, etc) & computers\\
\hline
\textbf{Probing} & Questions/answers & Questions/answers mediated & Behavioural evaluation\\
\textbf{mechanism} & mediated by natural & by physicochemical & (sensitivity to external\\
& language & language (chemical & stimuli, behavioural\\
& &  potentials, mechanical,  & differences, etc.)\\
&  &  transduction, signalling, etc.) & mediated by a lossless\\
&  & & compression algorithm.\\
\hline
   \end{tabular}
   \end{center}
\caption{\label{comptable} Comparison of Turing tests for intelligence, life \cite{cronin} and computation.}
\end{table}
\end{center}

The behavioural approach takes Turing's disembodied concept of universal computation independent of substratum to its logical limit, its central question being whether one can program a system to behave in a desired way. This is again close to Turing's test in which the interrogator cannot directly see the individual replying, because intelligence is not a property that requires the possessor to have a ``skin" (in the words of Turing himself \cite{turing}), for example, or to be a human being for that matter (Turing's approach), just as computation doesn't require electricity, or for that matter a digital computer (this approach). This approach that bases itself on the extent to which a system can be programmed tells us to what degree a given system resembles a computer. As the interrogator we will use a lossless compression algorithm that manifests properties of an observer, such as some type of subjectivity and finite resources. As suggested by Sutner \cite{sutner}, it is reasonable to require that any definition of computation in the general sense, rather than being a purely logical description (e.g. in terms of recursion theory), should capture some sense of what a physical computation might be. Sutner adds ``A physical system is not intrinsically a computer, rather it is necessary to interpret certain features of the physical system as representing a computation." This obliges Sutner to take into consideration the observer and the act of interpretation of a physical system.

 In many ways, this account of computation can be derived from the negation of Piccinini's 4th. feature (\emph{the wrong things do not compute}) \cite{piccinini}, which I think is dogmatic and gets in the way of the extension of the notion of computation to cover natural computation. Among the things that Piccinini rules out as objects that possibly compute are planetary systems, hurricanes and digestive systems. In fact, Piccinini himself seems to have some difficulty (\cite{piccinini}, p. 508) justifying how a digestive system is not computational. For insofar as a legitimate mechanistic account can be given of a digestive system, that would mean that it possesses precisely the sorts of properties and components that are taken into consideration in determining whether or not a system counts as a computer. I will argue that one doesn't need to axiomatically rule out such systems as computing or not. I will avoid making claims about whether or not such systems compute, because the approach advanced herein is above all a pragmatic approach designed to have applications (in fact it was first developed as a tool for the investigation of dynamical properties of computer programs and not primarily as a philosophical account). 

On the other hand, the behavioural account defended herein does satisfy Piccinini's 3rd requirement (\emph{the right things compute}). Piccinini's requirements 2 (\emph{Explanation}) and 6 (\emph{Taxonomy}) are at the core of this proposal connecting programmability and computation and providing a grading system based on behaviour. Piccinini's requirement 5 (\emph{Miscomputation}) doesn't seem very relevant to this proposal, and even if it were, to this author this feature doesn't seem essential to computation, for it is hard to see how a computational system can miscompute other than in the eyes of the observer. Indeed Piccinini himself sees this as troublesome in an account of computation, as it violates requirement 1. In fact, weak (i.e. observer dependent) miscomputation is pervasive in nature; I think nature amply manifests this kind of ``miscomputation". In summary, I reject requirement 1 (the basis of Piccinini's account), satisfy requirements 2, 3, and 6, particularly 2 and 6 at which I think this proposal excels. And concerning requirement 4, I remain neutral, not to say unconvinced, although I can acknowledge a form of \emph{weak miscomputation}, that is a computation that does not go in the way the observer expects it to. This approach allows a taxonomy of computation.

\subsection{Algorithmic complexity as an approximative measure of programmability}
\label{measure}

The traditional connection between behaviour and computation has tended toward explaining behaviour as computation~\cite{hodgkin} or computation as emulating brain activity~\cite{mccullock}, but this author has no knowledge of explorations in the direction of explaining computation as behaviour.

 This paper proposes an alternative behavioural definition of computation based on whether a system is capable of reacting to the environment---the input---as reflected in a measure of \emph{programmability}. This will be done by using a phase transition coefficient previously defined in an attempt to characterise the evolution of cellular automata and other systems. This transition coefficient measures the sensitivity of a system to external stimuli and will be used to define the susceptibility of a system to being (efficiently) programmed, in the context of a nature-like definition of computation.

 Turing's observer is replaced by a lossless compression algorithm, which has subjective qualities just like a regular observer, in that it can only partially ``see" regularities in data, there being no perfectly effective compression algorithm in existence. The compression algorithm will look at the evolution of a system and determine, by means of feeding the system with different initial conditions (which is analogous to questioning it), whether it reacts to external stimuli.

 The compressed version of the evolution of a system is an approximation of its algorithmic (Kolmogorov) complexity defined by \cite{kolmo,chaitin}:

\begin{center}
$K_T(s) = \min \{|p|, T(p)=s\}$
\end{center}

That is, the length of the shortest program $p$ that outputs the string $s$ running on a universal Turing machine $T$)~\cite{kolmo,chaitin}. A technical inconvenience of $K$ as a function taking $s$ to be the length of the shortest program that produces $s$ is its non-computability, proven by reduction to the halting problem. In other words, there is no program which takes a string $s$ as input and produces the integer $K(s)$ as output. This is usually taken to be a major problem, but one would expect a universal measure of complexity to have such a property. The measure was first conceived to define randomness and is today the accepted objective mathematical measure of complexity, among other reasons because it has been proven to be mathematically robust (in that it represents the convergence of several independent definitions). The mathematical theory of randomness has proven that properties of random objects can be captured by non-computable measures. One can, for example, approach $K$ using lossless compression algorithms that detect regularities in order to compress data. The value of the compressibility method is that the compression of a string as an approximation to $K$ is a sufficient test of non-randomness. If the shortest program producing $s$ is larger than $|s|$ the length of $s$, then $s$ is considered to be random.

Based on the principles of algorithmic complexity, one can use the result of the compression algorithms applied to the evolution of a system to characterise the behaviour of the system~\cite{zenilca} by comparing it to its uncompressed evolution. If the evolution is too random, the compressed version won't be much shorter than the length of the original evolution itself. It is clear that one can characterise systems by their behaviour \cite{zenilca}: if they are compressible they are simple, otherwise they are complex (random-looking). The approach can be taken further and used to detect phase transitions, as shown in~\cite{zenilca}, for one can detect differences between the compressed versions of the behaviour of a system for different initial configurations. This second measure allows us to characterise systems by their sensitivity to the environment: the more sensitive the greater the variation in length of the compressed evolutions. A classification places at the top systems that can be considered to be both efficient information carriers and highly programmable, given that they react succinctly to input perturbations. Systems that are too perturbable, however, do not show phase transitions and are grouped as inefficient information carriers. The efficiency requirement is to avoid what is known as Turing tarpits~\cite{perlis}, that is, systems that are capable of universal computation but are actually very hard to program. This means that there is a difference between what can be achieved in principle and the practical ability of a system to perform a task. This approach is therefore sensitive to the practicalities of programming a system rather than to its potential theoretical capability of being programmed. What if, instead of trying to draw a crystal clear line between what is and is not a computer, one were to define a measure of (\emph{``computedness"})? I propose the following approach as a first approximation to \emph{programmability}. 

Let $C$ be an approximation to $K$ (given that $K$ is non-computable) by any means, for example, by using lossless compression algorithms or using the coding theorem technique we presented in \cite{delahayezenil}. Let's define the function $f$ as the variability of a system $M$ as the result of fitting a curve $\phi$ (by (linear) regression analysis) to the data points produced by different runs of increasing time $t^\prime$ (for fixed $n$) up to a given time $t$, of the sums of the differences in length of the approximations to Kolmogorov complexity ($C$) of a system $M$ for inputs $i_j$, $j\in\{1, \ldots, n\} \in E$, divided by $t(n-1)$ (for the sole purpose of \emph{normalising} the measure by the system's ``volume," so that one can roughly compare different systems for different $n$ and different $t$). With $E$ an enumeration of initial inputs for $M$. The following expression is a more formal attempt to capture t  compressed lengths of $M$ for different initial conditions $i_j$. $M_t(i)$ is a system $M$ running for time $t$ and initial input configuration $i$. At the limit $\mathbb{C}_t^n$ captures the behaviour of $M_t$ for $t \rightarrow \infty$, but the value of $\mathbb{C}_t^n$ depends on the choices of $t$ and $n$ (we may sometimes refer to $\mathbb{C}$ as assuming a certain $t$ and $n$), so one can only aim to capture some average or asymptotic behaviour, if any (because no convergence is guaranteed). $\mathbb{C}_t^n$ is, however, an indicator of the degree of programmability of a system $M$ relative to its external stimuli (input $i$). The larger the derivative, the greater the variation in $M$, and hence in the possibility of programming $M$ to perform a task or transmit information at a rate captured by $\mathbb{C}_t^n$ itself (that is, whether for a small set of initial configurations $M$ produces a single significant change or does so incrementally).  Now the second step is to define the asymptotic measure, that is the derivative of $f$ with respect to time, as a system's programmability (first basic definition):

\begin{equation}
\label{index}
\centering
\mathbb{C}_t^n(M)= \frac{\partial f(M,t,n)}{\partial t}
\end{equation}


For example, as is shown in \cite{zenilca}, certain elementary cellular automata rules that are highly sensitive to initial conditions and present phase transitions which dramatically change their qualitative behaviour when starting from different initial configurations can be characterised by these qualitative properties. A further investigation of the relation between this transition coefficient and the computational capabilities of certain known (Turing) universal machines has been undertaken in \cite{zeniluniversalca}. We will refrain from exact evaluations of $\mathbb{C}$ to avoid distracting the reader with numerical approximations that may detract from our particular goal in this paper. Other calculations have been advanced in  \cite{zenilpt} and \cite{zeniljetai}.

\subsection{A behavioural approach to computation}

The following are first approaches to definitions connected to the qualitative behaviour of \emph{computational} systems:\\

\textit{Approximate variability (the number of possible different evolutions of a system):} Let $U_1, U_2, \ldots $ be an enumeration of inputs to a system $M$. We are interested in the question of how different the evolution of $M(U_i)$ is to the evolution of $M(U_j)$, in particular the maximum difference.\\

\textit{Programmability:} The capability of a system to change, to react to external stimuli (input) in order to alter its behaviour. Programmability, then, is a combination of variability and external control.\\

\textit{Computational universality:} Maximum programmability.\\

\textit{Efficient programmability:} Maximum variability changes reached in polynomial time (of a \emph{small} degree).\\

\textit{Efficient universal computation:} Universality with measurable variations detected in polynomial time (of a \emph{small} degree).\\

Notice how close this approach is to Turing's test for intelligence. This is a kind of generalisation of the Turing test: \emph{computation is what behaves as such}, and it does so if it can be programmed.

 The following assertions follow (a technical paper with formal definitions is in preparation): 

\begin{itemize}
\item A system $U$ is capable of computation if $\mathbb{C}_t^n(U)>0$ for $t,n>0$. 
\item A 0-computer is not a computer in any intuitive sense because it is not capable of carrying out any calculation. \item A system capable of (Turing) universal computation has a non-zero $\mathbb{C}$ limit value (see \cite{zeniluniversalca}). (A non-zero $\mathbb{C}$ value, however, doesn't imply Turing universality.) 
\item A system $U$ capable of Turing computational universality asymptotically converges to $\lim \mathbb{C}_t^n(U) = 1$ for $t,n \rightarrow \infty$.
\end{itemize}



\begin{figure}[htdp]
\label{turingtest1}
\centering
  \scalebox{.25}{\includegraphics{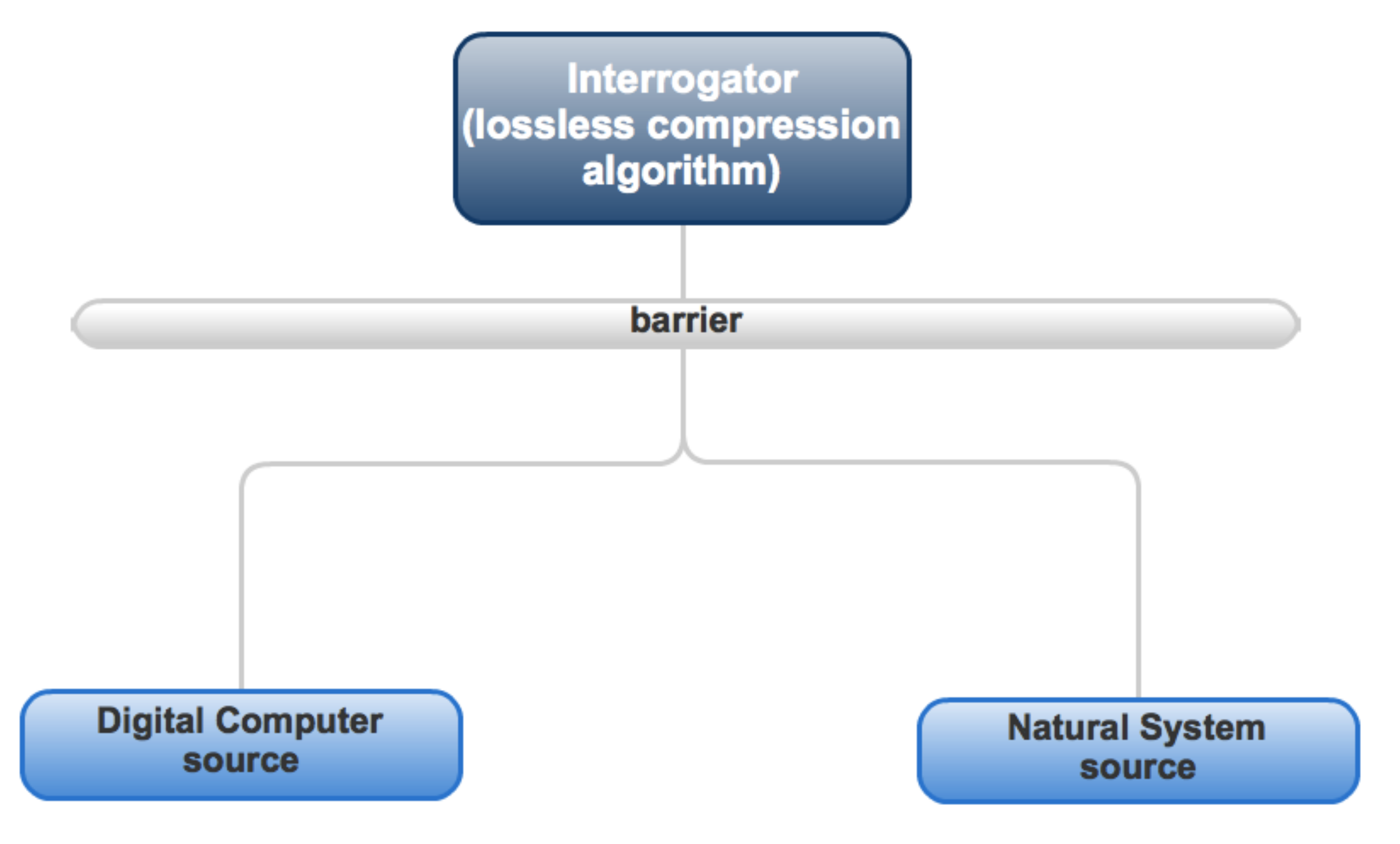}}\\
\caption{The Turing-test inspired approach to the question of computation as a behavioural test undertaken by a lossless compression algorithm in the role of the answer evaluator. Notice that the natural system can be a human being or anything else.}
\end{figure}

The use of a general lossless compression algorithm is comparable with the role of an interrogator in Turing's test (see Fig.~\ref{turingtest1}). To the compression algorithm the carrier of the computation is irrelevant as long as it can be represented in some form such that it can serve as input when running said compression algorithm. On the other hand, a compression algorithm is resource bound, in that it cannot implement in a finite time all the tests that can effectively detect all possible regularities in the data. This means that the algorithm is somehow subjective; it will first resort to what strikes it as the most obvious  patterns to use to compress the data. Yet the algorithm does this in a sophisticated way, with a greater likelihood of success than a human compressor, as it is systematic and implements general methods. Lossless compression algorithms can also be set to run for a longer time to attempt more methods of compression, just as a human observer would devise more methods of compression given more time.

So a system $S$ is provided with a random input $i$ (a ``question") and the lossless compression algorithm evaluates the reaction of the system (then mapping the input $i$ to a numerical value $C(S(i))$, the compressed length of $S(i)$ using the compression algorithm $C$). Just as observers would do for regularity appreciations (or answer evaluations), different compression algorithms may retrieve different compression lengths of $S(i)$, as they may differ in the way they compress. This compressed value is not completely arbitrary, as there is some objectivity in a strong desirable sense. This is because lossless compression is a sufficient test of non-randomness, meaning that if a lossless compression algorithm $C$ is able to compress $S(i)$ then the Kolmogorov complexity of $K(S(i))$ cannot be greater than $(C(S(i)))$. On the other hand, no $C^\prime$ algorithm can compress $S(i)$ such that $(C(S(i)))<K(S)$ by definition of $K$, so the values of a compression algorithm $C$ are not completely arbitrary (or subjective).

One may challenge the configuration depicted in~\ref{turingtest1} as lacking a true questioner, given that the compression algorithm evaluates the answers but does not formulate the questions, meaning that the test, unlike Turing's, is not self-contained. This is a very good and legitimate point, but thanks to Turing, it is not very well founded. This is because from Turing we know that a system $S$ with input $i$ can be rewritten as a new system $S^\prime(\langle S\rangle,i)$, that is a new system $S^\prime$ encoding $S$ with input $i$. One can actually do this not just for a single input, but for any number of inputs, even an infinite number of inputs such as in an enumeration. Let $E$ be an enumeration for $S$ and $p_E$ the program that produces $E$ (we know that the program exists by definition). Then $S^\prime(\langle S\rangle,\langle p_E\rangle)$ such that $S^\prime$ behaves like $S$ and uses $p_E$ to feed $S$ with an infinite number of inputs (just as $S$ for $i$, $S^\prime$ may not halt). So in some strong sense the system is neutral even to having all the questions at once or not.

\section{A taxonomy of computation}
\label{taxonomy}

The measure proposed in~\ref{measure} can be used to dynamically define computation based on the \emph{degree of programmability} of a system. The advantage of using the transition coefficient $\mathbb{C}$ is that it is indifferent to the internal states, formalism or architecture of a computer or computing model; it doesn't even specify whether a machine has to be digital or analog, or what its maximal computational power must be. It is only based on the behaviour of the system in question. It allows us to minimally characterise the concept of computation on the basis of behaviour alone. 

Now we can attribute the property of computation to natural and physical objects, hence arriving at a measure of \emph{Nature-like computation}, and distinguish between the computational attributes of physical objects depending on their programmability. 

Our proposal has many similarities to Piccinini's mechanistic approach, yielding a hierarchy of computing objects. But while he puts calculators and (specific-purpose) computers in different categories, I don't see any essential reason to do so. He places the concept of programmability at the centre of the discussion, as I do, but all in all our approaches are very different. His mechanistic approach doesn't seem particularly suitable for natural computation. At a more fundamental level, Piccinini's approach differs from this approach in that he seems to attribute importance to the physical implementation of a computation and to its physical components, whereas this is not a matter of interest here. Unlike Piccinini, I do not think that the property of computing is an objective feature of a system.

 A program can be defined as that which turns a general-purpose computer into a special-purpose computer. This is not a strange definition, since in the context of computer science a computation can be regarded as the evolution undergone by a system when running a program. However, while interesting in itself, and not without a certain affinity with our approach, this route through the definition of a general-purpose computer is a circuitous one to take to define computation. For it commits one to defining computational universality before one can proceed to define something more basic, something which ideally should not depend on such a powerful (and even more difficult-to-define) concept. Universal computation is without a doubt the most important feature of computation, but every time one attempts to define computation in relation to universal computation, one ends up with a circular statement [computation is (Turing) universal computation], thus merely leading to a version of a CT thesis.

 As Piccinini suggests in \cite{piccinini}, a Turing universal computer, and indeed a human being, can do more than follow one algorithm. They can follow any algorithm, which is typically given to them in the form of instructions. ``More generally, a human can be instructed to perform the same activity (e.g. knitting or playing the piano) in many different ways. Any machine that can be easily modiÞed to yield different output patterns may be called `programmable'. In other words, `being programmable' means being modiÞable so as to perform relatively long sequences of different operations in a different way depending on the modiÞcation."

 If everyday things like fridges or lamps can be deemed computational, then it's hard to see how any physical system whatsoever is not computational (this relates to Putnam's realisation theorem, see Subsection~\ref{implementation}). We can now meaningfully ask the question whether a lamp or a fridge is or isn't a computer, without trivialising the question itself or any possible answer. A lamp's output, for example, can be described by two different behaviours (in this case, traditionally identified as states), that is, on and off, which are triggered by external input (via a switch). Even if the lamp can be considered to react to external stimuli, it is very limited in its behaviour, and the space of its initial configurations is finite and small (it has only two possible initial configurations). Hence the slope of the differences of the behavioural evolution in time is very close to 0. A lamp is therefore a very limited computer with $\mathbb{C}$ value very close to 0. If one wished to rule out lamps or fridges as computing devices one would only need to define a threshold beyond which a system can be said to compute and beneath which it would not be said to compute. With a definition of programmability one can expect to be able to construct a \emph{hierarchy of \emph{computing} objects} (see Table~\ref{hierarchy}), with digital general-purpose computers placed correctly (at the top of the hierarchy of computers), while other objects that we may consider (non) computing objects can be found at or near the bottom. It is clear that the threshold is at the level of specific-purpose computers, given that we may want to include in the definition of computation entities that compute simple functions such as---only---the successor function, or the sum of 2 integers, while we may not be able to assign any computing capabilities to a specific-purpose ``computer" capable of---only---``computing" the identity function.

\begin{center}
\begin{table}[h]
   \begin{center}
      \tabcolsep=0.12cm
   \begin{tabular}{|c|c|c|}
\hline
 \textit{Object} & $\mathbb{C}$ \textit{value} & \textit{Threshold flag} \\
& & ($\mathbb{C}>\delta$?)\\
\hline
General-purpose digital&&\\
(electronic) computer & $\mathbb{C}>>\delta>0$ & Yes\\
\hline
Human brains & $\mathbb{C}>>\delta>0$ & Yes\\
\hline
Specific-purpose computers&&\\
(e.g. calculators, successor machine) & $\mathbb{C}\geq0<\delta$ & Yes/No\\
\hline
Lamps & $\mathbb{C}\sim0<\delta$ & No\\
\hline
Rocks & $\mathbb{C}=0<\delta$ & No\\
\hline
   \end{tabular}
   \end{center}
\caption{\label{hierarchy} A primitive hierarchical view of computation according to the first approximation of computation based on the coefficient $\mathbb{C}$ with customisable threshold $\delta$ is considered a computer if $\mathbb{C}>\delta$, otherwise it is not. The symbol ``$>>$" is for systems for which (assuming they operate as they usually do, e.g. a fully capable human brain) no mistake about their computational capabilities can be made based on their degree of programmability approached by $\mathbb{C}$. That is, their $\mathbb{C}_t^n$ value is strictly greater than $\delta$ for any $\delta$ for $t$ and $n$ that are run for long enough (that is, long enough to be greater than $\delta$).}
\end{table}
\end{center}

\begin{figure}[htdp]
\label{rule255}
\centering
  \scalebox{.36}{\includegraphics{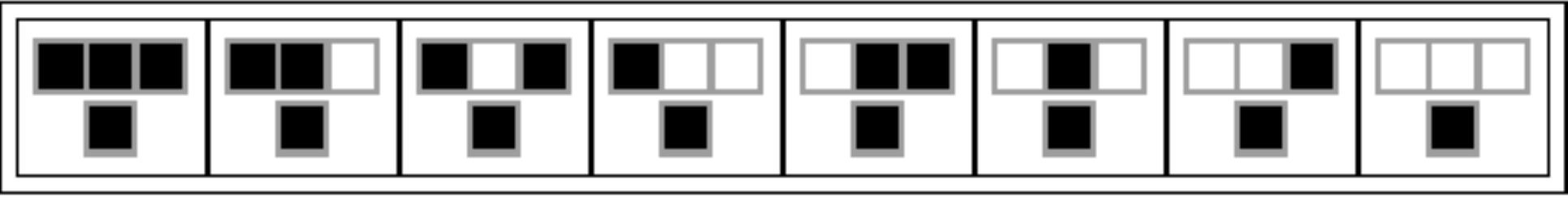}}\\
  \medskip
    \scalebox{.54}{\includegraphics{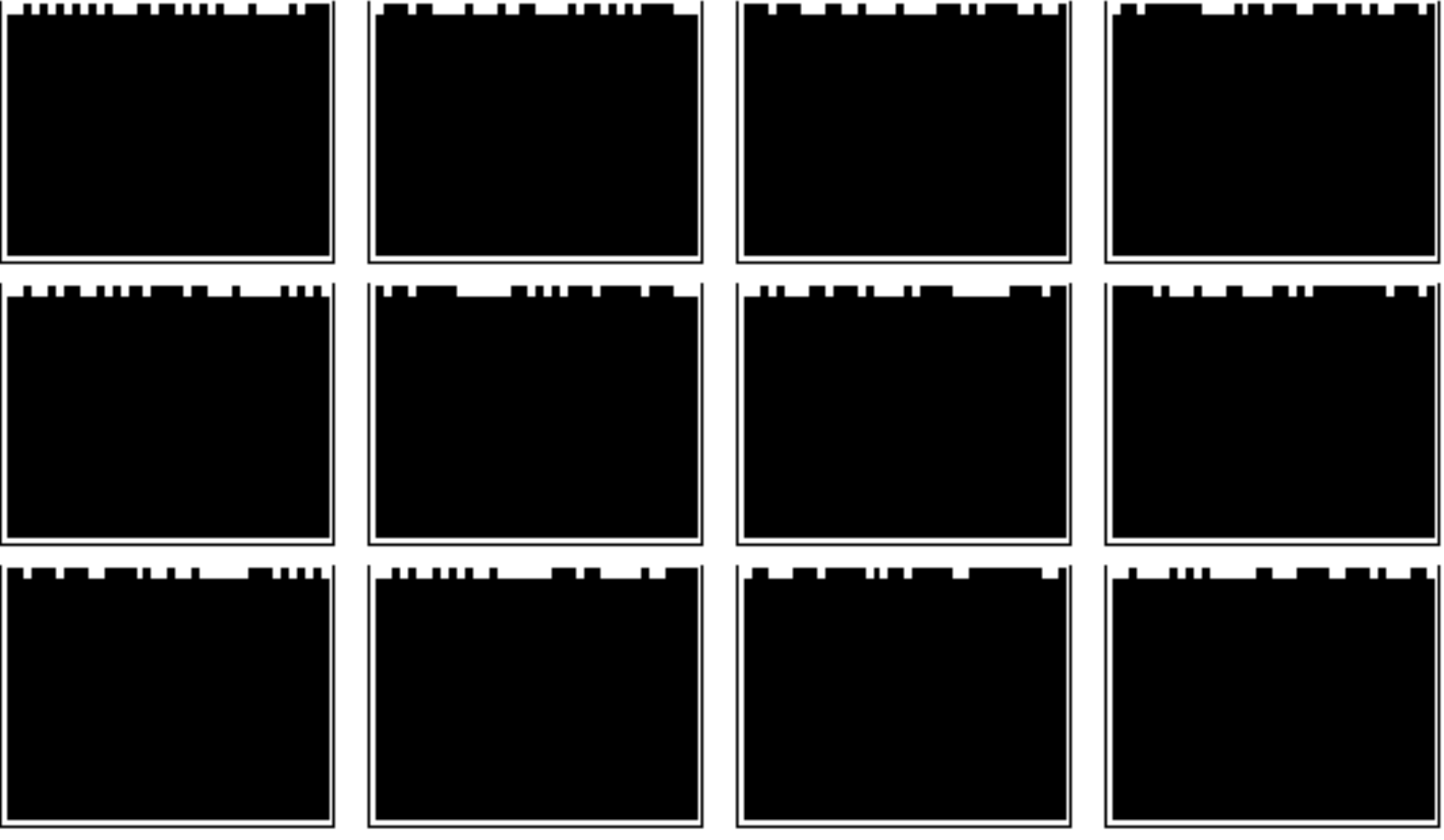}}\\
\caption{Example of a ``rock-like" behaviour by an elementary cellular automaton \cite{wolfram} with Rule 4 for several ``random" initial configurations and evolving from top to bottom. The evolutions are preceded by an icon (top) illustrating the rule that the system follows for every possible cell configuration.}
\end{figure}

Brains and digital (Turing universal) computers can show great variation for two different random inputs, potentially even for two arbitrarily close inputs (according to same sensible distance of inputs), but for systems with low $\mathbb{C}$ this is different. For example, a lamp has only two possible ``random" inputs: \emph{on} and \emph{off}, and the same number of outputs. For rock-like systems (including rocks themselves), the rock looks the same disregarding the possibly thinkable inputs for a rock. Fig.~\ref{rule255} shows a rock-like behaviour of an elementary cellular automaton.


According to Piccinini, we distinguish computers from most other things because, at the very least, computers are more versatile than other computing mechanisms. He thus attributes a measure of positive versatility to the concept of computation (or the computer). ``Computers can do arithmetic but also graphics, word processing, Internet browsing, and a myriad other things", Piccinini says. And he adds: ``Computer versatility calls for an explanation" \cite{piccinini}. Some objects, such as abaci, have parts that need to be moved by hand. They may be called computing aids, as Piccinini does. Of course abaci would have very small, if not zero, $\mathbb{C}$ values with no human intervention, and therefore can be flagged as non-computers, even for small $\delta$ threshold value. 

This account does justice to digital computers and the practices of computer scientists and computability theorists. On the one hand, digital computers, calculators, both universal and nonuniversal Turing machines, and finite state automata, are examples of computation under the proposed definition. These can be recognised as computers, and universal digital computers can be placed at the top of the hierarchy of computational systems. On the other hand, the definition also places the concept of \emph{programmability} at the centre of the practice of computer science, but through algorithmic complexity one can also define higher classes of computation based on Turing degrees, given that abstract machines that can solve the halting problem behaviourally perform a computation that cannot be carried out by a Turing machine that may not halt. That is, the oracle machine does halt, but it does not halt for every possible computation; it has its own new halting problem of a higher degree, and so on, hence building up the arithmetical hierarchy without need of explicit descriptions of states or functions. 

It is clear that computers are not the only programmable mechanisms. So are brains, as are many other natural systems that we can now control and direct to perform certain tasks that they were not supposed to be naturally capable of (e.g. through genetic engineering). A computer is a system that can be modified to compute in different ways. I think one, if not the most important features of brains and computers is that they can be reprogrammed (in different but analogous ways). Along the lines of Fodor's conclusion~\cite{fodor}, but with no need of state representation, if the brain is a programmable system, then it is a computer under this behavioural approach. This is paradoxical because according to Fodor's slogan ``no computation without representation," according to which, in order for a system to be ascribed computational status, it needs to be construed as representing information in some way. According to the view proposed here, we should neither reject or accept this dictum because a natural occurring process can be assigned a computational value if and only if it can be programmed, regardless of whether it represents anything. As pointed out by Dresner~\cite{dresner}, a measurement-theoretic representation typically is accompanied by a uniqueness theorem that states how all the homomorphisms from the given empirical structure to the numerical one relates to each other (that is, can be obtained from each other). I will provide some clues of how to do this in the answer to possible objections in Section~\ref{objections}.

Beyond formalisms, the present account of computers and computation is used to formulate a rigorous taxonomy. According to this behavioural approach, all Turing machines that compute a function other than identity are computers, and all that do so are universal Turing machines.  It encompasses minds and computers while excluding almost everything else, investing minds and computers with a special status. One may think of some possible counterexamples. Think of the billiard ball computing model. It is designed to perform as a computer and can therefore be trivially mapped onto the states of a digital computer. Yet it is a counterexample of what the semantic account sets out to do, viz. to cordon off minds and computers (believed capable of computation) from things like billiard balls, tables and rocks (believed to be incapable of computation). The billiard ball computing model, as a system, however, is identified as computational in this behavioural approach, without further ado.






\section{Addressing possible objections}
\label{objections}

Despite avoiding representationalism, which is one advantage of this approach, we find that certain objections to Turing's test, including some addressed by Turing himself, can serve as objections to the behavioural approach to computation, and that possible objections to the behavioural approach to computation can also serve as objections to the Turing test. Nevertheless, we claim that the behavioural approach can provide useful tools for natural computation, and we will use it as a basis for a set of measures capturing different properties of the dynamic behaviour of natural systems, measures drawing on concepts from algorithmic information theory and compressibility. The objections are not thoroughly addressed here, as each may require a paper of its own, but I sketch some possible responses to explore.

\subsection{Technical objections}

Let me first address some possible technical objections before turning to the philosophical ones. These and other objections deserve careful scrutiny, but there is no reason to address them all in depth here. 


\subsubsection{The assumption that compressibility can capture different behaviours}

One assumption that the first approach to a definition of programmability makes is that compression algorithms are able to distinguish between different behaviours. From the proposed definition we derive the differences in the compressed lengths of the evolution of a system. But it may not be clear whether the length of the compressed version for a given initial configuration of a system can differ from the length of the compressed version of the evolution of a system for a different initial configuration that yields apparently different behaviour. The problem can be stated as follows: Imagine that one has two very different processes generating different data files, but that the lengths of their compressed versions using, for example, gzip, are the same. It may seem that our approach is suggesting that both processes are behaviourally the same, even when, apart from the coincidence in the compressed lengths of their respective outputs, they may in fact be completely different. 



$K(s)$, however, is \emph{upper semi-computable}~\cite{li}; there is a sequence of lossless compression algorithms approximating $K(s)$: $C_1(s) \geq C_2(s) \geq C_3(s) \geq \ldots \geq K(s)$. That is, one can find a sequence of compression algorithms that asymptotically approaches $K$. $K(s)$ cannot then be greater than the most compressed version of $s$. The invariance theorem~\cite{chaitin} in the theory of algorithmic information guarantees that the outputs can be distinguished from one another at the limit, no matter how close they are to each other, by a compression algorithm approaching $K$, and up to a bounded degree of precision (which can be large, but increasing $t$ eventually overcomes it). More formally, the invariance theorem states that if $C_U(s)$ and $C_{U^\prime}(s)$ are the shortest programs generating s using the universal Turing machines $U$ and $U^\prime$ respectively, their difference will be bounded by an additive constant independent of $s$. 

It is easy to see that the underlying concept is that since both $U$ and $U^\prime$ are universal Turing machines, one can always write a general translator (a compiler) between $U$ and $U^\prime$ such that one can run either Turing machine and get one or another complexity value, simply adding the constant length of the translator to the result.

 This means that eventually, if two processes are essentially different in the sense of algorithmic complexity, they will have different $C$ values from some time $t$ on up to $K$. The caveat that a system may be characterised in an imprecise fashion still applies, but the invariance theorem guarantees that the approach is sound theoretically, even if in practice it may sometimes be misleading, in a way that we are used to with compression algorithms that may not ``see" regularities in a file (e.g. a file containing the digits of $\pi$). 

It is worth noticing that two different evolutions produced by the same rule system, such as a cellular automaton, may not necessarily have the same Kolmogorov complexity (in fact it is unlikely they will if they appear different) because the system in question is $S(i)$ and not $S$ alone, that is $S$ for the initial configuration $i$ (e.g. Rule 30 elementary cellular automaton~\cite{wolfram} starting from a black cell is a different system than Rule 30 starting from a repetition of ten times 01). From Turing's universality, we know that $S(i)$ can always be rewritten as $S^\prime$, that is a system with empty input that behaves like $S$ for input $i$, where it is clear that $S \neq S^\prime$, and this difference is ultimately captured by the difference between $K(S)$ and $K(S^\prime)$, that is the lengths of the shortest programs producing $S$ and $S^\prime$.





\subsubsection{The choice of enumeration of initial configurations}
\label{enumeration}

The interrogator plays an important part in this Turing-based approach, which is why the initial input configurations are key---their role is analogous to that of the interrogator questioning the system. In general, one can always tamper with an enumeration $E$ to make a system behave in a certain way for a limited period of time, as one can always run a system and then pick initial conditions for which the system behaves in a certain way, proceeding to design another enumeration $E^\prime$ for which the first $E^\prime_t$ members are members of $E$ but sorted from $t=1, ..., n$ such that the system behaves in a desired way for the first $n$ elements. So how sensitive to the choice of initial input enumeration is the Turing-test inspired approach to the problem of natural computation? One can make $n$ as large as one wishes, but the limit behaviour of a system will always go beyond $n$. Does this guarantee that from some point on (e.g. $n$) the system will start behaving ``naturally"? Imagine that one knew that a system behaved in a certain way for even length initial configurations. One could then design a $E$ such that all initial configurations are of even length. But $E$ has to reach every possible initial configuration in finite time, so there is no way to design $E$ so that it would run all even length inputs and then all odd length inputs in a finite time. There is no way to fool the limit analysis of the behaviour of a system by tampering with the initial configurations for more than a finite number of inputs.



The general question of the appropriate enumeration of inputs for a system is worth exploring, especially for natural systems, given that it is not always clear what the enumeration of inputs for a natural system might be (questions arise, for example, about continuous-value parameters that may need to be discretised in order for a compression algorithm to analyse). One obvious problem is that of ``encrypted systems". What if an efficient programmable computing system looks intentionally random and inefficient? Say one Turing universal system (e.g. Rule 110~\cite{wolfram,cook}) behaving like another random-looking system (such as Rule 30 in the same rulespace). It is still Rule 110, but the question is whether one would be able to identify and program Rule 110 if it is behaving like Rule 30. It may be that one can only know it is Rule 110 if one knows the decrypting function, so the compression algorithm can be fooled. This is related to who can pass a ``stupidity test", that is a system that is so smart that it
knows how to look stupid, or to really is (one cannot pass, however, an intelligence or computation test without being intelligent or being able to compute. ). The question of ``encrypted systems" occurring in nature is important to address. But this is certainly related to a feature I think is desirable in this behavioural approach, that of observer-dependent subjectivity (Subsection~\ref{observer}) and to the question of the enumeration of initial conditions (Subsection~\ref{enumeration}) and the question of some sort of minimal need for representation (Subsection~\ref{representation}).

\subsection{Foundational objections}

It is interesting to see how some objections serve at once as arguments against the Turing machine intelligence approach and this natural computational approach, while others do not (e.g. the Mathematical Objection (Searle~\cite{searle2}, Penrose~\cite{penrose}) doesn't seem obviously to apply to the question of computation). Other examples are the theological and the consciousness arguments, which work against both machine intelligence and natural computation by endowing  humans and natural things with \emph{qualia}, which are said not to be concomitants of the domain of digital computation. The objections work differently however, because in the case of machine intelligence they are meant to ``safeguard" the essence of the human being, endowing it with irreproducible qualities such as consciousness, while in the case of natural computation they work to ``safeguard" the nature of digital computation. The advantage of my approach as compared to Turing's is that there are fewer people willing to defend machines than humans, though heated debates are carried out in both directions. The current tendency in computation, however, is greater openness to the possibility that objects and systems other than electronic computers compute. 


\subsubsection{Some representation is needed}
\label{representation}

\begin{figure}[htdp]
\label{turingtest2}
\centering
  \scalebox{.25}{\includegraphics{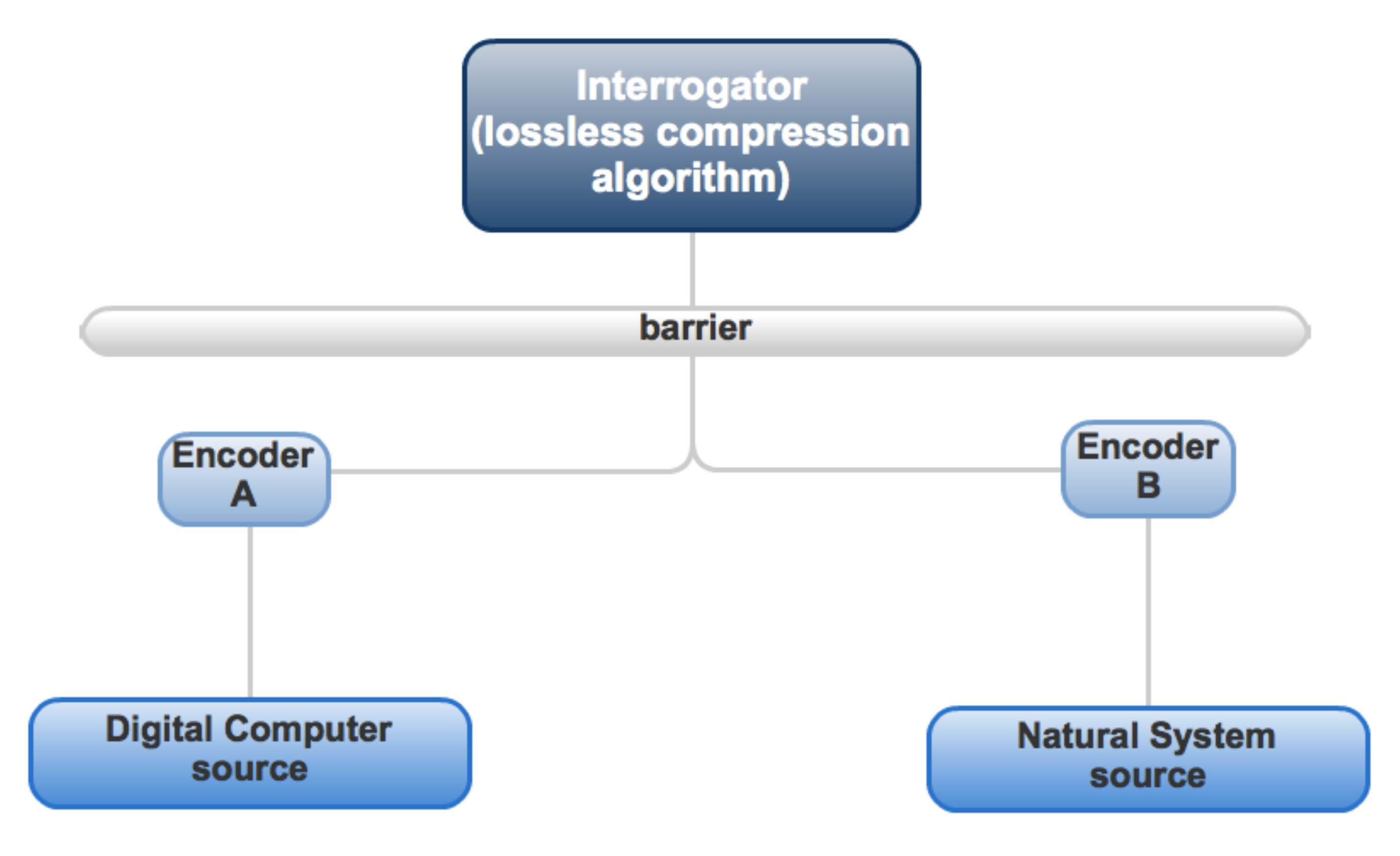}}\\
\caption{What is the nature of the encoders? They work in both directions encoding ``questions" properly for each system, and feeding the lossless compression algorithm in the right format. Encoders A and B may be of very different nature. Simple encoders always seem a possibility, but questions about their implementation and role remain legitimate.}
\end{figure}

It is interesting to note that one needs some representation of the output of a system before feeding the compression algorithm (see Fig.~\ref{turingtest2}). What about the introduction or the simplification of complexity in the encoding process from the language of the system to the language of a digital computer implementing the lossless compression algorithm?

 This is indeed the case, and it implies that there is some communication and mapping between the natural system and the digital computer implementing the lossless compression algorithm, but this mapping is of a very different nature from the mapping of states or functions among systems. Is this representation always possible? 

On the one hand, one can always discretise data. On the other hand we know that a discrete language can always be translated into binary. So in a technical sense this is always possible. This is related to the previous discussion of the question of whether a universal system could emulate a random-looking system to hide its programmability capabilities, and what this would mean.

The proposal advanced herein is, however, different to the requirement of a strong form of representationalism, where knowing the states of a system to put them into relation with another is needed, which is fully dependent on the complete (and unlikely) knowledge of the states of a natural system. Here, however, it is only needed a weak translation of one system output language into another, represented by the encoders boxes in Fig.~\ref{turingtest2}. Encoders should not be seen as a drawback, we deal with them all the time in computing in the form of compilers. In systems development different programming languages are used for different purposes in different places of a system. Even if one can simulate each other as Turing complete, very few non-trivial applications today are fully developed in a single programming language.

\subsubsection{Programmability}

A second immediate reaction is whether placing programmability at the centre of a definition of computation is too strong as an assumption. For one may think of artificial and natural systems that may not look \emph{programmable}, yet one would be ready to grant can compute (e.g. discrete neural networks). I think this objection arises from a conflation between the standard meaning of programming and the behavioural one I am advancing here. While it is true that for many artificial and natural systems a concept of programmability is difficult to determine, the concept of programmability advanced in this paper is about whether one can, by any means, make a system behave in a way other than the way it was already behaving. In this sense, for example, a logic circuit or a batch process may not qualify as a computation if these are unable to react to external stimuli or if the observer is unable to witness such an interaction if it happens in the design or the launch of a computing process.

\subsubsection{Human-machine and intentionality objections}
\label{intentionality}

When Julien Offray de La Mettrie \cite{mettrie} took Descartes' method to what he claimed was its logical conclusion in his L'homme-machine, the argument was that Descartes' attempt to defend the theory of a human soul by relegating mechanical behaviour to animals in fact acted against humans. For if animals were capable of feeding, moving and interacting with other animals, strictly speaking, there was nothing to prevent human behaviour from being seen as a consequence of mechanical behaviour. In the Turing test we see a similar reversal of the argument, where it is not the machine's intelligence that is questioned but rather the intelligence of the human being, not because the questioners harbour the suspicion that humans may not be intelligent but because the mechanisms that drive human intelligence may turn out to be of the same order as those that drive computers today. 

Searle advances the problem of intrinsic meaning or ``intentionality" \cite{searle2}. Harnad \cite{harnad} defines it as the symbol grounding problem. I consider this objection weak in our context (though it remains to be further explored), because if assumed, the definition of computation is rendered meaningless in the physical context (we know we can program certain natural things, these things would be considered computers when computing for us, and not when not). For the Turing test, some ``intention" is desirable, as Turing is trying to convince his reader that there is no argument in principle for a machine to fail an intelligence test if it increasingly improves its performance when imitating intelligent human behaviour. Also it is clear that electronic computers, back in Turing's time as well as today, are assembled for the purpose of computing, hence no harm is done by assuming some intentionality. 

Dennett has suggested \cite{dennett} that it would seem that explicit representation is not necessary for the explanation of propositional attitudes. For example, during a game of chess with a computer program, attitudes such as  ``It thinks that the queen should be moved to the left" are often attributed to the computer. Yet no one would suggest that the computer actually thinks or believes, in the way we do. I think it is clear how this behavioural approach to computation is compatible with this view, and neutral on intentionality questions, as it is only interested in the ways a system seems to behave and not whether it ``really" does so (meaning it intended to do so, whether we are concerned with computers or with natural systems, including the brain).

\subsubsection{The observer-oriented objection}
\label{observer}

One immediate reaction to, and a possible objection to this approach, concerns the applicability of such a behavioural (observer-oriented) definition, given the possibly arbitrary choice of $\delta$ (see e.g. Table~\ref{hierarchy}). According to certain arguments, computation is observer-relative, either in the sense that many physical systems implement many computations (Putnam \cite{putnam}), or in the sense that almost all physical systems implement all computations (Searle \cite{searle}). Some physical objects, for example, may be seen to implement any computation of whatever complexity. Thus the walls in Searle's \cite{searle}, implement his wordprocessing program. Since the physical description of an object underdetermines its computational description in this way, computation is deemed observer-relative \cite{searle}. 

This is of course a legitimate objection, which also applies to other behavioural approaches to other notions, such as the notion of intelligence, and to the Turing test. I have suggested (Section~\ref{measure}), however, that a measure of limit behaviour is possible, and that even if $\delta$ is very large, one can always overcome it over time for systems that are indubitably computing devices according to the programmability approach (computers and brains), while one can always contrive to have trivial devices such as lamps and rocks not pass as computers, leaving a flexible space in between for systems that may or may not, subjectively, be considered computers.

Its dependence on programming language or universal Turing machine has traditionally been considered one of the drawbacks of Kolmogorov complexity. In this approach we actually take advantage of this property of Kolmogorov complexity, as it assorts with a behavioural approach to computation that cannot but be observer-(or machine-) relative. This is because the Kolmogorov (program-size or algorithmic) complexity only makes sense once a universal Turing machine or Turing-complete language is fixed. On the other hand, because Kolmogorov complexity ($K$) is uncomputable (another commonly identified drawback), or more precisely, upper semicomputable, it is what the compression algorithm ``observes" that approaches $K$ that we will turn to our advantage in capturing the qualitative behaviour of a computational system in order to quantitatively measure it. 

Piccinini argues that any reasonable definition of computation should be objective. I don't think, however, that this should be a sine qua non of a reasonable account of computation, nor that failure to meet this objectivity criterion makes an account vacuous or trivial. In fact I think computation is intrinsically user/observer oriented, both in practice and in theory. In practice, computation is mostly, if not entirely about programming systems. On the one hand, programming systems is intentional (driven by the desire to make a computer behave in a particular way), even if intentionality is not essential to computation. On the other hand, theory prescribes subjectivity in various ways. The halting problem can be read as an observer-relative property of computational systems, given that one cannot, in general, ever know whether a computation will halt except by running a system for a number of steps---which depends on the willingness of the observer to wait, if it doesn't halt before the specified number of steps. The problem is not exclusive to halting, but extends to reachability in general, that is, the question of whether a system will reach a certain configuration. Universal computation is subjective in the sense that one has to decide to stop a computation and deal with the fact that one may never know whether such a computation will ever halt or reach a certain configuration.

\subsubsection{The halting problem prescribes subjectivity at all levels}

The halting problem is the problem of deciding whether a computation will halt or not. The halting problem implies that computations can be divided into 2 categories: reducible and irreducible, that is computations that are simple enough to be determined to halt or never halt, and computations for which the only option is to run them and wait for them to halt, which may obviously take an infinite amount of time. Irreducible computations can also be classified into 2 kind of computations: computations that never halt and therefore not even running the computation will help and computations that halt in time $t$ but there is no way to know $t$ but by running the computation for at least $t$. Clearly this characterisation incorporates an important role of the observer in that there exists computations for which one
can only know whether they will halt by running them, and introduces a subjective component, namely the fact that the observer has to decide a runtime cutoff that is willing to wait before making an informed assumption about the (non-)halting characteristic of a computation.

Now one can see how an observer is important in the account of computation even for the most classical case of the unsolvability of the halting problem. This is even more evident when considering other phenomena, such as reachability, that is whether a computation will reach certain configuration, in which for some computations only an observer willing to run and witness the computation may answer.

The undecidability of the halting problem affects all theoretical and practical notions related to computation. For example, in Kolmogorov complexity one can never say whether an object is random (one can say whether an object is simple if it has been compressed but not the converse). This doesn't make algorithmic complexity useless. In fact it is this observer-relative property (with respect to compression algorithms that may or may not ``see" regularities in the data) that the measure is most useful---for all kinds of applications, including classification of animal species and languages by compressibility, detection of genetic sequences, fraud and plagiarism detection. In finite Kolmogorov complexity, finite randomness is in the eye of the beholder, in the sense that any finite sequence can always be part of a random or non-random string. Hence the quality of being random is observer-dependent, just as it is in the case of the halting problem. 

I think that it is denying the role of the observer that makes the intuitive notion of computation vacuous or trivial. The observer plays an essential role in the definition of computation. This is made explicit in our definition of computation, for the purposes of  generalising and characterising natural computation. 

Under this approach, computation is observer-relative (in agreement with many authors who endorse computationalism), just as intelligence is observer-relative in the case of the Turing test. We find that certain objections to Turing's test, including some addressed by Turing himself, can serve as objections to this behavioural approach to computation (we will address some of them), and that possible objections to the behavioural approach to computation can also serve as objections to the Turing test. Nevertheless, we claim that the behavioural computation approach can provide useful tools for natural computation, and we will use it as the basis for a toolkit of quantitative measures---based on concepts from algorithmic information theory and compressibility---capturing different qualitative properties of natural systems. 

Paradoxically, the behavioural approach does not explain a system's behaviour, at least not in full, for we can explain part of a system's behaviour once a first behavioural analysis is performed, but not in the way we would be led to expect if we followed Smith or Piccinini, for it is not intended to be a theory of computing, nor does it set out to fully account for the causes of a system's behaviour, only for its apparent behaviour. The approach proves to have applicability and to provide insight into the properties of dynamical systems about whose internal states one could potentially have no information, nor any clue as to the possible mappings between a natural and an abstract computational system. But it also works well for systems which we know and whose internal states we can study in full detail, producing all manner of mappings to other models of computation, as we have shown using cellular automata and the way in which the measures based on this behavioural approach allow us to characterise phase transitions or rates of information transfer from a purely behavioural perspective. 

Take the example of having to calculate the Lyapunov exponents of a natural system. Even if the system can be described as a dynamical system for which orbits can be described, this already assumes that one is able to represent such dynamics. Of course the behavioural approach also assumes that one can capture the behaviour of the system, but it does not assume full knowledge of the precise evolution of the system. In fact one can to some extent analyse a system in an instant of time without having to go through intermediate times (this will of course impact the final result, as it improves in direct proportion the more one observes the system). 

If the observer is essential to the definition of computation, one has to acknowledge that there is no sense to the most general question of whether the universe computes, because no definition of the universe allows for external stimuli (external to the universe), nor for the output of the universe to reside outside it for an observer to evaluate.

\subsubsection{Does implementation matter?}
\label{implementation}

The question of the implementation of computation seems not to have been taken seriously until critics of computationalism brought forward certain arguments to the effect that a great many physical systems implement many, if not all, computations. Such arguments have been presented by Putnam \cite{putnam} and Searle \cite{searle}. According to Putnam (the eponymous Putnam's Realization Theorem), ``for every ordinary open system $S$, for every finite state automaton $M$ (without input and output), for any number $n$ of computational steps of the automaton $M$, and for every real-time interval $I$ (divisible into $n$ subintervals) $S$ realizes $n$ computational steps of $M$ within $I$". And according to Searle (what is sometimes called Searle's Thesis), ``for any program and for any sufficiently complex (physical) object, there is some description of the object under which it is implementing the program." 

Along the lines of the question asked by Chalmers \cite{chalmers}, what makes a rock compute something (or nothing) rather than everything? It seems that, at least prima facie, what (abstract) computability and (concrete) computation have in common is some logical description, only the characterisation of the latter isn't exhausted by a purely logical description, so implementation does matter. And it does matter in my approach, given that the rock may potentially be capable of any computation (think of using its particles to build a more programmable device), but it does not do so at the level at which it must be  described as a rock, and if we look at it through a Turing-test inspired lens and attempt to make it behave in one way or another, i.e.  program it to behave differently for different external stimuli (see Fig.~\ref{rule255} for a ``rock-like" behaviour of an abstract system).

\subsubsection{Laws have no distinguished character}

It has been suggested \cite{parsons} that I am assigning a special status to physical laws, or to computer programs for that matter. This is an understandable objection but in fact it represents a misconception of my position. The misunderstanding resides in the conclusion that by connecting laws to computer programs as opposed to data, I give physical laws a special, immortal and unchanging status. Computer programs, however, can be written in bits. And, as I have explained in Section~\ref{universality}, Turing proved that computer programs and data are not essentially different; one can always exchange one for the other. That is, it is possible to write the transition table of a Turing machine in the form of an input for a universal Turing machine, or to build a transition table (a Turing machine) from the computer program description. 

In algorithmic probability there is only one strong assumption regarding the distribution of objects. What Levin's universal distribution is supposed to indicate is the probability of a string being generated by a program, but one has to make an assumption as regards the distribution of programs in order to talk about \emph{picking a random program}. And that is the only possible uninformed assumption--the uniform distribution. That is, any program of the same length is equally likely to occur as a product of chance. But apart from this one is free to interchange programs. There is nothing special about physical laws. They can be seen as highlighting or summarising a regularity in the data (the world), and data can change, hence physical laws may do so as well.

\subsubsection{The question of scale}

In the real world, things are constituted by smaller elements unless they are elementary particles. One therefore has to study the behaviour of a system at a given scale and not at all possible scales, otherwise the question becomes meaningless, as elements of a physical object are molecules, and ultimately atoms and particles that have their own behaviour, about which too the question about computation can be asked. This means that a $\mathbb{C}$-computer may have a low or null $\mathbb{C}$ at some scale but contain $\mathbb{C}^\prime$-computers with $\mathbb{C}^\prime > \mathbb{C}$ at another scale (for which the original object is no longer the same as a whole). A setup in which $\mathbb{C}^\prime \leq \mathbb{C}$ is actually often common at some scale for any computational device. For example, a digital computer is made of simpler components, each of which at some macroscopic level but independently of the interconnected computer, is of lower behavioural richness and may qualify for a $\mathbb{C}$ of lower value. In other words, the behavioural definition is not additive in the sense that a $\mathbb{C}$-computer can contain or be contained in another $\mathbb{C}^\prime$-computer such that $\mathbb{C} \neq \mathbb{C}^\prime$. 

In the physical world, under this qualitative approach, things may compute or not depending on the scale at which they are studied. To say that a table computes only makes sense at the scale of the table, and as a $\mathbb{C}$-computer it would have a very limited $\mathbb{C}$, that is a very limited range of behaviour, for it can hardly be programmed to do something else.

The behavioural definition is not immune to scale. Something may or may not compute at a certain level of description but it may compute at another more macro- or more microscopic level of description. But the concept of the object is also not scale invariant (we call things by different names when we change scale, e.g. we call the constituents of a rock atoms, or the aggregation of $H_2O$ in liquid form water).

\subsubsection{Batch process objection}

A batch process is the execution of a program on a computer without the need of any external intervention. This kind of system would go unnoticed by this proposed behavioural approach given the insensitivity of such a system to any external stimuli, as it is programmed to perform a task without interacting with anything else until it stops and produces some output, if any. During this time the process may look as if it were doing nothing, but this is merely appearance, and there are ways for the observer to ascertain that it is in fact computing, at the lowest level by its external resource consumption and release, such as energy and heat (which one could also manipulate to make the process change behaviour, for example, stop the process), and at another level, by monitoring the process for a long-enough time. The batch process instance is only valid as an objection between the time $t=1$ when the process is actually initiated (it has to), and $t=n-1$, because at least at one time $t=0$ or $t=n$ (if it halts and produces an output) some interaction with the outside is expected to happen. So while some computers may fail to be identified by the behavioural definition, the limit behaviour definition seems to be immune to this objection, except insofar as it may for all (proper) purposes consider something that may be computing as not computing because it is disconnected from the external world in which the observer lives.

\subsubsection{The contingency of quantum mechanics}

Using algorithmic probability (AP) S. Lloyd claims \cite{lloyd}:
\begin{quote}
I would suggest, merely as a metaphor here, but also as the basis for a scientific program to investigate the computational capacity of the universe, that this is also a reasonable explanation for \emph{why} \emph{the universe is complex. It gets programmed by little random} \emph{quantum fluctuations}, like the same sorts of quantum fluctuations that mean that our galaxy is here rather than somewhere else.
\flushright
(S. Lloyd, 2002)
\end{quote}

We don't know whether AP can be adapted to a quantum version but we do know that there is no need for \emph{quantum fluctuations} to generate algorithmic structure \cite{delahayezenil} that Lloyd was trying to explain on the basis of quantum mechanics.

The strong assumption in the context of classical computation and classical mechanics is \textit{determinism}. The wave-function collapse in quantum mechanics and the problem of measurement may challenge determinism at two different levels, but otherwise classical mechanics prescribes determinism in the (macroscopic) universe. Classical (Newtonian) mechanics guarantees the deterministic output (the problem is to generate the same input). Running a computation twice with the same input generates the same output through exactly the same path just as would do a classical system following the rules of classical mechanics (that in practice this is not possible is due to the problem of limited accuracy of the measurement of the initial conditions).

\subsubsection{Connections to computational complexity}

In the light of this research now one can find an interesting connection of the measure $\mathbb{C}$ to traditional computational complexity where one is concerned with the needed resources for a computation to be carried out. $\mathbb{C}$ provides clues on whether a system may be Turing universal but not on whether a system may not be universal, because universality requires variability and sensitivity to external stimuli to program a computation. Also $\mathbb{C}$ is greatly influenced but not directly related to universality given that universality will guarantee that $\lim{\mathbb{C}_t^n = \infty}$ for $t,n \rightarrow \infty$, but a positive value $\mathbb{C}$ does not guarantee universality, it guarantees sensibility which in this context is a measure of the capability of the system to be programmed to do different (even if limited) computations by transferring information from the input to the output. But $\mathbb{C}$ ultimately depends on the way in which $\mathbb{C}$ is calculated for a finite number of initial configurations and a finite number of steps, hence systems that may compute at a slow pace may be misclassified for some $t$ and $n$ small enough.  $\mathbb{C}$  can be, however, thought as also measuring efficiency of a system to be programmed. So one can relativise this concept introducing time complexity classes. So one can say that a system with  $\mathbb{C}$ value that grows in linear time is efficient, but it is not efficient if it grows in logarithmic time.

\section{Concluding remarks}

This paper has addressed the problem of recognising computation. It partially fulfils some of the requirements that according to several authors any definition of computation should meet (e.g. \cite{scott}, \cite{piccinini}), while I have made the case that some properties are not needed and should not be required or expected, especially in the novel context of natural computation and artificial biology.

Computational models can be very useful even when not every detail about a system is known. The aim of systems biology, for example, is to understand the functional properties and behaviour of living organisms, while the aim of synthetic biology is to design, control and program the behaviour of living systems, even without knowing the details of the biological systems in question. Along the lines of Turing's intelligence test, this approach seems to be useful for investigating qualitative properties of computing systems in a quantitative fashion, and since it places programmability at the centre of computation it serves as a possible foundation for natural computation.

\section*{Acknowledgements}

I would like to thank Marcin Mi\l kowski for some references and the organisers of the symposium \textit{Natural/Unconventional Computing and its Philosophical Significance} for their kind invitation to speak at the AISB/IACAP World Congress 2012---Alan Turing 2012. I also wish to thank the FQXi for the mini-grant awarded by way of the Silicon Valley Foundation under the title ``Time and Computation", which this project studies in connection to behaviour (mini-grant no. 2011-93849 (4661)).

\bibliographystyle{AISB}

\begin{thebibliography}{00}

\bibitem{auslander} Ausl\"ander, S., Ausl\"ander, D., M\"uller, M., Wieland, M. and Fussenegger, M. Programmable single-cell mammalian biocomputers, \emph{Nature}, 2012.
\bibitem{gol}  Berlekamp, E.R., Conway, J.H., Guy, R.K. \emph{Winning Ways for your Mathematical Plays,} AK Peters Ltd., 2001.
\bibitem{blanco} Invited talk by Blanco, J. \emph{Interdisciplinary Workshop with Javier Blanco: Ontological, Epistemological and Methodological Aspects of Computer Science}, University of Stuttgart, Germany, July 7th 2011.
\bibitem{chaitin} Chaitin, G.J. On the length of programs for computing finite binary sequences: Statistical considerations, \emph{Journal of the ACM}, 16(1):145--159, 1969.
\bibitem{chalmers} Chalmers, D.J. Does a Rock Implement Every Finite-State Automaton? \emph{Synthese}, 108:310--333, 1996.
\bibitem{cook} Cook, M. Universality in Elementary Cellular Automata, \emph{Complex Systems} 15: pp. 1--40, 2004.
\bibitem{conrad} Conrad, M. The Price of Programmability, In R. Herken (ed) \emph{The Universal Turing Machine, A Half-Century Survey,} Springer-Verlag, 1994.
\bibitem{conway} Berlekamp, E., Conway, K. and Guy R. Winning Ways for your Mathematical Plays, vol. 2, Academic Press, 1982.
\bibitem{cronin} Cronin, L., Krasnogor, N., Davis, B.G., Alexander, C., Robertson, N., Steinke, J.H.G., Schroeder, S.L.M., Khlobystov, A.N., Cooper, G., Gardner, P.M., Siepmann, P., Whitaker, B.J., and Marsh, D. The imitation game---a computational chemical approach to recognizing life, \emph{Nature Biotechnology}, vol. 24, no. 10, 2006.
\bibitem{davis} Invited talk by Davis, M. Universality is Ubiquitous, Invited Lecture, \emph{History and Philosophy of Computing (HAPOC11),} Ghent, 8 November, 2011.
\bibitem{dennett} Dennett, D.C., \emph{Brainstorms: Philosophical Essays on Mind and Psychology}, MIT Press, 1981.
\bibitem{delahayezenil} Delahaye, J.-P. and Zenil, H., Numerical Evaluation of the Complexity of Short Strings: A Glance Into the Innermost Structure of Algorithmic Randomness, \emph{Applied Math. and Comp.}, 219, pp. 63-77, 2012.
\bibitem{dresner} Dresner E., Measurement-theoretic representation and computation-theoretic realisation, \emph{The Journal of Philosophy,} vol. cvii, no. 6, 2010.
\bibitem{fodor} Fodor, J. \emph{The Language of Thought,} Harvard University Press, 1975
\bibitem{fredkin} Fredkin, Ed. Finite Nature, \emph{Proceedings of the XXVIIth Rencotre de Moriond}, 1992.
\bibitem{harnad} Harnad, S. The Symbol Grounding Problem, \emph{Physica D}, 42: 335--346, 1990.
\bibitem{hodgkin} Hodgkin A. and Huxley, A. A quantitative description of membrane current and its application to conduction and excitation in nerve, \emph{Journal of Physiology}, 117:500--544, 1952.
\bibitem{kolmo} Kolmogorov, A.N. Three approaches to the quantitative definition of information, \emph{Problems of Information and Transmission}, 1(1):1--7, 1965.
\bibitem{langton} Langton, C.G. Studying artificial life with cellular automata, \emph{Physica D: Nonlinear Phenomena} 22 (1-3): 120--149, 1986.
\bibitem{lloyd} Lloyd, S. Computational capacity of the Universe, \emph{Physical Review Letters}, 88, 237901, 2002.
\bibitem{li} Li, M. and Vit\'anyi, P.  \textit{An Introduction to Kolmogorov Complexity and Its Applications.} Springer, 3rd., 2008.
\bibitem{mccullock} McCullock, W. and Pitts, W. A Logical Calculus of Ideas Immanent in Nervous Activity, \emph{Bulletin of Mathematical Biophysics} 5 (4): 115--133, 1943.
\bibitem{minsky} Minsky, M. \emph{Computation: Finite and Infinite Machines,} Prentice Hall, 1967.
\bibitem{parsons} Parsons, W.T. Patterns in the Noise: Physics as the Ultimate Environmental Science, \emph{FQXi's 2012 Essay Contest,} ``Questioning the Foundations".
\bibitem{penrose} Penrose, R. \emph{The Emperor's New Mind}, Oxford University Press, 1989.
\bibitem{perlis} Perlis, A.J. Epigrams on Programming, \emph{SIGPLAN Notices,} Vol. 17, No. 9, pages 7--13, 1982.
\bibitem{piccinini} Piccinini, G. Computers, \emph{PaciÞc Philosophical Quarterly,} 89, 32--73, 2008.
\bibitem{putnam} Putnam, H. \emph{Representation and Reality,} Cambridge: MIT Press, 1988.
\bibitem{margolus} Margolus, N. Physics-like Models of Computation, Physica, Vol. 10D, pp. 81--95, 1984.
\bibitem{scott} Scott, D.S. Outline of a mathematical theory of computation, \emph{Technical Monograph PRG-2, Oxford University Computing Laboratory,} England, November 1970.
\emph{Theoretical Computer Science,} Volume 412, pp. 183--190, 2011.
\bibitem{searle} Searle, J.R. \emph{The Rediscovery of the Mind,} MIT Press, 1992.
\bibitem{searle2} Searle, J.R. Is the Brain a Digital Computer, in \emph{Philosophy in a New Century,} pp 86--106, Cambridge University Press, 2008.
\bibitem{sutner} Sutner, K. Computational Processes, Observers and Turing Incompleteness, \bibitem{turner} Turner, R. Specification, \emph{Minds and Machines,} 21 (2): pp 135--152, 2011.
\bibitem{mettrie} Thomson, A. \emph{Machine man and other writings}, Cambridge University Press, 1996.
\bibitem{turing2} Turing, A.M. Systems of logic based on ordinals, \emph{Proc. London math. soc.}, 45, 1939.
\bibitem{turing} Turing, A.M. Computing Machinery and Intelligence, \emph{Computing machinery and intelligence,} Mind, 59, 433--460, 1950.
\bibitem{wolfram} Wolfram, S. \emph{A New Kind of Science,} Wolfram Media, 2002.
\bibitem{woods} Woods, D. Neary, T., The complexity of small universal Turing machines: a survey, \emph{Theor. Comput. Sci.} 410(4-5): 443--450, 2009.
\bibitem{zenilca} Zenil, H. Compression-based investigation of the behaviour of cellular automata and other systems, \emph{Complex Systems,} (19)2, 2010.
\bibitem{zenilminds} Zenil, H., Soler-Toscano F. and Joosten, J.J. Empirical Encounters With Computational Irreducibility and Unpredictability, \emph{Minds and Machines,} vol. 22, Number 3, pp. 149-165, 2012.
\bibitem{zeniluniversalca} Zenil, H. On the Dynamic Qualitative Behaviour of Universal Computation, \emph{Complex Systems,} (20)3, 2012.
\bibitem{zenilpt} Zenil, H. What is Nature-like Computation? A Behavioural Approach and a Notion of Programmability, \emph{Philosophy \& Technology}, 2012, DOI: 10.1007/s13347-012-0095-2.
\bibitem{zeniljetai} Zenil, H. Programmability for Natural Computation and the Game of Life as a Case Study, \emph{J. of Experimental \& Theoretical Artificial Intelligence}, forthcoming.

\end{thebibliography}

\end{document}